\documentclass[11pt]{article}
\usepackage[a4paper,hmarginratio=1:1,vmarginratio=2:3,totalwidth=15.6cm,
totalheight=21.cm]{geometry}
\usepackage{bm,epstopdf,epsfig,amsmath,amssymb,amsfonts,colordvi,
latexsym,comment,cancel,verbatim}
\usepackage[font=md,captionskip=8pt]{subfig}
\usepackage[usenames,dvipsnames]{color}

\renewcommand{\baselinestretch}{1.13}

\newcommand{\cD}{{\cal D}}

\newcommand{\cJ}{{\cal J}}

\newcommand{\cO}{{\cal O}}

\newcommand{\ra}{\rightarrow}
\newcommand{\be}{\begin{equation}}
\newcommand{\ee}{\end{equation}}
\newcommand{\bea}{\begin{eqnarray}}
\newcommand{\eea}{\end{eqnarray}}
\newcommand{\Ra}{\Rightarrow}

\newcommand{\otheta}{\overline\theta}
\newcommand{\opsi}{\overline\psi}

\long\def\symbolfootnote[#1]#2{\begingroup
\def\thefootnote{\fnsymbol{footnote}}\footnote[#1]{#2}\endgroup} 
\begin{document}

\begin{flushright}
CERN-PH-TH/2011-219\\
CPHT-RR067.1011
\end{flushright}

\thispagestyle{empty}

\vspace{2.2cm}

\begin{center}
{\Large {\bf Goldstino and sgoldstino in microscopic models

\bigskip
and the constrained superfields formalism.
}}

\bigskip
\vspace{1.cm}
\textbf{
I. Antoniadis$^{\,a}$\,\symbolfootnote[2]{on leave from CPHT (UMR CNRS 7644) 
Ecole Polytechnique, F-91128 Palaiseau, France.}, 
E.~Dudas$^{\,b,\,c}$,
D.~M. Ghilencea$^{\,a,\,b,\,d}$
\symbolfootnote[3]{E-mail:
ignatios.antoniadis@cern.ch, dumitru.ghilencea@cern.ch,
Emilian.Dudas@cpht.polytechnique.fr}} \\

\vspace{0.3cm}
$^a$ {\small CERN - Theory Division, CH-1211 Geneva 23, Switzerland.}\\
$^b$ {\small CPHT (UMR CNRS 7644) Ecole Polytechnique, F-91128 Palaiseau, France.}\\
$^c$ {\small LPT  (UMR CNRS 8627), Bat 210,
 Universite de Paris-Sud, F-91405 Orsay Cedex, 
France.}\\
$^d$ {\small  DFT, National Institute of Physics 
and Nuclear Engineering (IFIN-HH) Bucharest MG-6, Romania.}

\end{center}

\def\baselinestretch{1.14}
\begin{abstract}
\noindent
We examine the exact relation between the superconformal symmetry breaking 
chiral superfield  ($X$) and the goldstino superfield in microscopic models
of an arbitrary Kahler potential ($K$) and in the presence of matter fields.  
We  investigate the decoupling of the massive sgoldstino 
and scalar matter fields  and the offshell/onshell-SUSY  expressions of
their superfields in terms of the fermions composites. 
For general $K$ of two superfields, we  study the properties 
of the superfield $X$ after integrating out these scalar fields, 
to show that in the  infrared  it satisfies (offshell) the condition $X^3=0$ 
and $X^2\not=0$. We then compare our results to those of the well-known method 
of constrained  superfields discussed in the literature,  
based on the  conjecture $X^2=0$. Our results can be used in 
applications, to couple offshell  the (s)goldstino fields to  
realistic models such as the MSSM. 
\end{abstract}


\newpage

\section{Introduction}

A consequence of supersymmetry breaking is the presence of a Goldstone fermion
- the goldstino - and its scalar superpartner, the sgoldstino. In supergravity 
the goldstino is the longitudinal component of the gravitino which becomes massive
(of mass $\sim\! f/M_{Planck}$), while the sgoldstino field
acquires  mass and decouples at low energy, just like Standard 
Model (SM) superpartners.  If the scale of supersymmetry breaking  
($\sqrt f$) is low (an example is gauge mediation)
the goldstino couplings ($\sim\! 1/\sqrt f$) are much stronger than the
Planck-suppressed couplings of transverse gravitino and one can work in
the gravity-decoupled limit, with a massless goldstino.

The study of the interactions of the goldstino and sgoldstino with matter
is an interesting research area that 
started with the pioneering work of Akulov and Volkov \cite{Volkov:1973ix}
and were investigated extensively in the early days of supersymmetry 
\cite{cddfg}-\cite{Samuel:1982uh},
with renewed  interest in \cite{Casalbuoni:1988xh}-\cite{Dudas:2011kt}.
One way to describe  Goldstino interactions is using the  formalism
of constrained superfields which has a long history, that started with
\cite{Ivanov:1977my}, \cite{Rocek:1978nb}, \cite{Lindstrom:1979kq},  
\cite{Samuel:1982uh}, and more recently in  \cite{Casalbuoni:1988xh}.  
For a review of the constrained superfields 
method with new  insight and detailed examples, we refer the reader to \cite{SK}. 
The role of the constraints  applied to the matter and goldstino superfields is
to project out (i.e. integrate out) the heavy SM superpartners and sgoldstino that 
decouple at low energy. The method is useful to describe in a superfield 
formalism,  nonlinear effective Lagrangians obtained after 
integrating  out, in a Susy model, all heavy superpartners.

To learn more about the couplings of goldstino/sgoldstino to matter
fields one  starts from the conservation equation of the 
Ferrara-Zumino current $\cJ$ \cite{Ferrara} that has the form 
$\overline D^{\dot\alpha}\,\cJ_{\alpha\dot\alpha}=D_\alpha\,X$. 
$X$ is a chiral superfield that breaks supersymmetry and conformal symmetry and
 was only recently proposed to be identified \cite{SK} 
in the infrared (IR) limit, to the goldstino superfield.
By considering an IR constraint of the form $X^2=0$, one 
shows that in the IR  the sgoldstino becomes a composite of 
goldstino fields,  $\phi\propto \psi\psi/F$. This constraint, 
when added to a Lagrangian
$L=\int d^4\theta \, X^\dagger X + \big[ \int d^2\theta \, f\,X+h.c.\big]
+\cO(1/\Lambda)$
provides, onshell, a superfield description of the Akulov-Volkov Lagrangian 
for the goldstino \cite{Casalbuoni:1988xh,SK}. 
The constraint  $X^2\!=\!0$ brings interactions to the otherwise free theory of
$L$ and  projects out  the  sgoldstino. Such constrained $L$
can arise in the low energy limit from an effective Lagrangian with
additional non-renormalizable $\cO(1/\Lambda)$ corrections 
to $L$, that provide a large sgoldstino mass; this is then 
integrated out via the eqs of motion and generates the constraint
($\Lambda$ is a UV scale). This is the situation in the  absence of matter 
fields.

In the presence of matter (super)fields the situation is more subtle.
In general, if one would like to couple the goldstino (super)field to the MSSM,
this can be done via the divergence of the supersymmetric current and the 
equivalence  theorem \cite{cddfg}. Alternatively,  one can use
effective polynomial interactions between  the MSSM and 
goldstino superfields \cite{SK},  as detailed in \cite{adgt}. 
In the presence of matter fields, the superfield constraint 
that projects out the sgoldstino was  conjectured \cite{SK} 
to remain (in IR) if the form 
$X^2=0$, while the constraint that projects out the massive
 scalar component $\phi^j$ of 
a matter  superfield $\Phi_j$ can be taken $X\,\Phi_j=0$. 
These can be used to identify the sgoldstino and scalar matter 
fields after they decouple,
as functions (composites) of the (light) fermions present; for particular examples
see \cite{SK} but also  some counterexamples in \cite{Dudas:2011kt}. 
The onshell/offshell validity limits of these conjectured constraints in  the 
presence of matter fields are discussed in this paper in a general setup
of microscopic models with an arbitrary Kahler potential.

For a model independent approach to the goldstino physics, we use
the solution for $X$ of the Ferrara-Zumino  conservation equation, that can 
be found in \cite{Clark:1995bg}  for an arbitrary  Kahler $K$ and  superpotential $W$;
this shows that $X$ is a combination of all the fields  present. 
Surprisingly, the implications 
of this result for goldstino microscopic models were little studied
and in this work we explore this direction.  
Taking advantage of this result, we compute in a model independent
way for an arbitrary $K$,  the expression of $X$ after integrating out, 
via the eqs of motion,
the massive sgoldstino and scalar matter fields that decouple in IR. 
We prove that for two superfields case and a linear superpotential 
the field $X$ then satisfies (offshell) 
the condition $X^3=0$  while $X^2\not =0$ (for more superfields 
and additional interactions this can be generalized to higher order conditions).
This result is in disagreement with the conjecture $X^2=0$ mentioned earlier,
as the general condition to identify the sgoldstino field. 
Further, for an arbitrary  $K$, we evaluate the offshell/onshell 
sgoldstino  and the scalar matter fields  as composites of the light fermions
 (goldstino and  matter fermions).  We show that these scalar
fields have expressions such that their superfields  satisfy (offshell)
higher order polynomial  conditions, such as
$\Phi_1^3=\Phi_1^2\,\Phi_2=\Phi_1\,\Phi_2^2=\Phi_2^3=0$. 
This was also noticed previously in particular examples in  \cite{Dudas:2011kt}. 
We consider these conditions can be more fundamental than the property $X^3=0$ 
mentioned, which depends on assumptions about the so-called improvement term, 
due to which $X$ is not unique. 
Let us stress that while we also investigate various
onshell-Susy conditions for superfields, the constraints are really 
interesting and most relevant when valid offshell.
Finally, the non-linear goldstino superfield 
can be used to couple it offshell  to a non-linear Susy
realization of the SM (see Section~5 of \cite{Dudas:2011kt}). 

The plan of the paper is as follows. Section~\ref{XX} computes the 
component fields of the superfield $X$ for  arbitrary $K, W$. 
Section~\ref{onefield} reviews the case of one superfield
(goldstino) and its exact relation to $X$, using both the method 
 of eqs of motion and the constrained superfield method. Applications
to two specific models are also provided.
Section~\ref{twofields}  evaluates, in the presence of matter fields,
the properties of $X$ after integrating out the sgoldstino and  the scalar 
matter field. The expressions of these scalar fields in terms 
of the light fermions present are also computed and the properties of
the corresponding non-linear superfields investigated.
The results are illustrated  for the case of two specific models.
Comparison with the method of constrained superfields is performed in 
Section~\ref{limits} to examine the validity limits of the latter. 
Our conclusions are presented in Section~\ref{conc}.

\section{The Lagrangian and the superfield $X$ for general $K$, $W$.}\label{XX}

In this section we consider a general action of arbitrary Kahler
$K(\Phi^i,\Phi_j^\dagger)$  and superpotential $W(\Phi^i)$ and compute the 
components of the $X$ superfield \cite{Clark:1995bg}; $\Phi^i$, $i=1,2....$ 
denotes a goldstino  chiral 
superfield ($i=1$) and additional matter fields ($i>1$), 
with components $(\phi^i,\psi^i,F^i)$. The general action is, after a Taylor 
expansion in Grassmann variables:
\medskip
\bea\label{LKW}
L& =& \int d^4\theta\,\, K(\Phi^i,\Phi_j^\dagger)+
\bigg[\int d^2 \theta\,\, W(\Phi^i)+
\int d^2 \otheta\,\, W^\dagger(\Phi_i^\dagger)\bigg]
\nonumber\\
&=& \!\!\!
K_i^{\,\,\,j} \Big[\partial_\mu\phi^i\,\partial^\mu\phi_j^\dagger
+\frac{i}{2}\,\big( \psi^i\,\sigma^\mu\cD_\mu\opsi_j
-\cD_\mu \,\psi^i\sigma^\mu\opsi_j\big)
+ F^i\,F_j^\dagger\Big]
\nonumber\\[4pt]
&+&
\frac{1}{4}\,K_{ij}^{kl}\,\psi^i\psi^j\,\opsi_k\opsi_l
+\Big[\big(W_k - \frac{1}{2} \,K_k^{ij}\opsi_i\opsi_j\big) 
\,F^k - \frac{1}{2}\,W_{ij}\,\psi^i\psi^j
+h.c.\Big]
\label{generalKW}
\eea

\medskip\noindent
where we ignored a $(-1/4)\Box K$ in the rhs and a sum over
 repeated indices is understood.
We denoted $K_i\equiv \partial K/\partial\phi^i$,
$K^n\equiv \partial K/\partial\phi^\dagger_n$,
$K_i^n\equiv \partial^2 K/(\partial\phi^i\,
\partial \phi_n^\dagger)$,
$W_j=\partial W/\partial \phi^j$, $W^j=(W_j)^\dagger$, etc,
with $W=W(\phi^i)$, $K=K(\phi^i,\phi_j^\dagger)$. 
The derivatives acting on the fermionic fields are
\medskip
\bea
\cD_\mu \psi^l&\equiv&\partial_\mu\psi^l-\Gamma^l_{jk}\,(\partial_\mu\phi^j)\,\,\psi^k,
\qquad 
\Gamma^l_{jk}=(K^{-1})^l_m\,K_{jk}^m\nonumber\\[4pt]
\cD_\mu \opsi_l&\equiv&\partial_\mu\opsi_l-\Gamma_l^{jk}\,
(\partial_\mu\phi_j^\dagger)\,\,\opsi_k,
\qquad
\Gamma_l^{jk} =(K^{-1})^m_l\,K^{jk}_m
\eea

\medskip\noindent
Eq.(\ref{generalKW}) is the offshell form of the Lagrangian.
The eqs of motion for auxiliary fields 
\bea
F_m^\dagger&=&-(K^{-1})^i_m \,W_i +\frac{1}{2}\,\Gamma_m^{lj}\,\opsi_l\opsi_j
\nonumber\\
F^m&=& -(K^{-1})^m_i \,W^i +\frac{1}{2}\,\Gamma^m_{lj}\,\psi^l\psi^j
\eea
can be used to obtain the onshell form of $L$.  That leads to a scalar potential
\medskip
\bea
V=(K^{-1})^i_j\,W_i\,W^j
\eea

\medskip\noindent
that shall be used later on. After a Taylor expansion about the ground state,
one  finds the mass of the sgoldstino and additional scalar matter fields present.

Let us introduce a chiral superfield ($X$) that is a measure of 
superconformal symmetry breaking
\bea
\overline D^{\dot \alpha} \cJ_{\alpha\dot\alpha}
=D_\alpha X,\qquad\quad X\equiv(\phi_X,\psi_X,F_X)
\eea
Here $\cJ$ is the Ferrara-Zumino current \cite{Ferrara}; the component
$\psi_X$ is related to the supersymmetry current and $F_X$ to
 the energy-momentum tensor. For the general, non-normalizable action 
 in (\ref{LKW}),  $X$ has been calculated in  
\cite{Clark:1995bg} (see also \cite{SK} for a discussion) and  has the 
form\footnote{The derivation of this formula is using the eqs of motion 
\cite{Clark:1995bg}.  Here we take this formula as general, and consider
that it applies/can be ``continued'' offshell too.}
\medskip
\bea\label{eq9}
X=4\,W- \frac{1}{3} \,{\overline D}^2\,K - \frac{1}{2} \,{\overline D}^2 
Y^\dagger
\eea

\medskip\noindent
that is valid for arbitrary $K$ and $W$.
The last term is the so-called improvement term 
where $Y(\Phi^j)$ is a holomorphic function
related to a Kahler transformation $K\ra K+3/2(Y+Y^\dagger)$, 
that does not change the eqs of motion.
The component fields of $\overline D^2 Y^\dagger$ are
$\overline D^2 Y^\dagger=(-4 F_Y^\dagger; -4 i\partial\!\!\!\slash\overline\psi_Y; 
4 \Box\phi_Y^\dagger)$.
In principle with a carefully chosen $Y$ one could in principle
try to simplify the form\footnote{For example one 
could ``cancel'' the $W$ dependence, 
by choosing $Y^\dagger=(-1/2)\,(D^2/\Box)\,W$.
However, in this case the solution 
for $Y$  would be non-local and is unacceptable. See also discussion 
in Section~\ref{Xp}, \ref{anotherex}.} of $X$.
In fact the presence of the Y-dependent terms renders $X$ rather arbitrary.
In the following we ignore the  improvement term effect on $X$ 
and set $\overline D^2 Y^\dagger=0$, and return to this later in the text.
Note that $X$ has mass dimension 3. To obtain a component form of $X$
that is needed later on, we use
 \bea\label{DK}
 {\overline D}^2 K &=&
 (-4)\,\Big[ K^j F_j^\dagger -\frac{1}{2}\,K^{ij}\,\opsi_i\opsi_j\Big]
 \nonumber\\
 &+&
 (4\,i)\,\,\sqrt 2\,\theta\,
 \Big[\sigma^\mu \,\big(\,
 K^j\,\partial_\mu\opsi_j+K^{ij}\,\,\opsi_j\,\partial_\mu\phi^\dagger_i\,\big)
 +i\,\psi^k\,\big(\,K^j_k\,F_j^\dagger-\frac{1}{2}\,K_k^{ij}\,\opsi_i\opsi_j\,\big)\Big]
 \nonumber\\
 &+&
(-4)\,\,\theta\theta\,\,\,\Big[L_{W=0}
-\partial_\mu
\Big(
K^j\,\partial_\mu \phi_j^\dagger
-\frac{i}{2} K_i^j\psi^i\sigma^\mu\opsi_j\Big)
\Big]
\eea

\medskip\noindent
Here $L_{W=0}$ is $L$ of eq.(\ref{generalKW}) with $W$ and all its derivatives
set to 0. We then find from (\ref{eq9})
\medskip
\bea\label{X}
X&=&
4\,W(\phi^i)+\frac{4}{3}\,\Big[K^j\,F_j^\dagger -\frac{1}{2} K^{ij}\,\opsi_i\opsi_j\Big]
\nonumber\\
&+&
\sqrt 2\,\, \theta\,\,
\Big\{
4 \,W_i\,\,\psi^i-\frac{4\,i}{3}\,
\Big[\sigma^\mu \,\big(\,
K^j\,\partial_\mu\opsi_j+K^{ij}\,\,\opsi_j\,\partial_\mu\phi^\dagger_i\,\big)
+i\,\psi^k\,\big(\,K^j_k\,F_j^\dagger-\frac{1}{2}\,K_k^{ij}\,\opsi_i\opsi_j\,\big)\Big]\Big\}
\nonumber\\
&+&\,\,\,\,\,\theta\theta\,\,\,
\Big\{
4\,\Big[ W_i\,F^i - \frac{1}{2}\,W_{ij}\,\psi^i\psi^j\Big]
+
\frac{4}{3}\,\Big[ L_{W=0} -
\partial_\mu
\Big(
K^j\,\partial_\mu \phi_j^\dagger
-\frac{i}{2} K_i^j\psi^i\sigma^\mu\opsi_j\Big)
\Big]\Big\}
\eea

\medskip\noindent
From this one immediately identifies the field components $(\phi_X,\psi_X,F_X)$
of the  superfield $X$ for a general Lagrangian, and we shall use this information 
in the following sections.

In the infrared, it was recently noted that  $X$ ``flows'' to a chiral superfield 
that for a single field case satisfies $X^2=0$, leading to an Akulov-Volkov  
action  in superfields \cite{SK}. In calculations, to take the IR limit one should
effectively impose $X^2=\cO(1/\Lambda)+\cO(\partial_\mu)$,  where  
$\cO(\partial_\mu)$ are derivative terms 
and $\cO(1/\Lambda)$  are non-derivative, $\Lambda$-suppressed terms,
that all vanish in the infrared;
here $\Lambda$ is the UV scale that enters 
in (\ref{LKW}) 
to suppress the higher dimensional effective operators present; for example in 
$K_{ij}^{kl}\sim 1/\Lambda^2$, $K^{ij}\sim 1/\Lambda$, etc.
In the presence of additional fields, a constraint for $X$ would actually mean
a constraint for a combination of  these fields, see eq.(\ref{eq9}). 

In the remaining sections we study the relation between $X$ and the goldstino
superfield ($\Phi_1$),  for the case of one or more superfields present and
for arbitrary $K$. The properties of $X$ are also discussed.
We then analyze the relation with the constrained superfields  formalism.

\section{The case of one superfield: Goldstino ($\Phi$) and the $X$ field.}
\label{onefield}  

\subsection{General results.}

To begin with, consider that we have only one superfield, the goldstino
itself $\Phi=(\phi,\psi,F)$,  and no matter superfields. We would like 
to clarify for an  arbitrary $K$, the offshell and onshell link between $X$ and
$\Phi$, in  the IR limit of setting $\partial_\mu$ derivatives to 0.

The action considered is that of (\ref{LKW}) with one superfield only ($\Phi$) and 
below we simplify  the notation  into $K_\phi=\partial K/\partial\phi$,
$K^\phi=\partial K/\partial\phi^\dagger$, $W_\phi\equiv \partial W/\partial \phi$, 
etc. Since $\Phi$ is a  goldstino superfield, we assume in (\ref{LKW}) that
 $W=f\,\Phi$, so  goldstino $\psi$  is indeed massless;
$F\sim-f+\cdots$ breaks Susy, while sgoldstino $\phi$ can acquire a large mass via 
higher dimensional Kahler terms and then decouples in IR. This is 
possible by integrating out additional massive fields (see later).  
With these remarks, 
we identify from (\ref{X}) the components of  $X=(\phi_X, \psi_X,F_X)$:
\medskip
\bea\label{ro2}
\phi_X &=& 
4\, W+\frac{4}{3}\Big[ K^\phi \, F^\dagger-\frac{1}{2} K^{\phi\phi}\,
\opsi\opsi\Big]
\nonumber\\
\psi_X 
&= &
\psi\,\frac{\partial \phi_X}{\partial \phi}-\frac{4\,i}{3}\,\sigma^\mu \,
\big(K^\phi\,\partial_\mu\opsi +K^{\phi\phi}\,\opsi\,\partial_\mu\phi^\dagger\big)
=
\psi\,\frac{\partial \phi_X}{\partial \phi}
+\cO(\partial_\mu)
\nonumber\\
F_X
&=&
\frac{4}{3}\,\Big\{
K_\phi^\phi\,\Big[\partial_\mu\phi\,\partial^\mu \phi^\dagger
+\frac{i}{2}\,\big(\psi\,\sigma^\mu D_\mu\opsi
-D_\mu\,\psi\,\sigma^\mu \opsi\big)\Big]
-
\partial_\mu \,\big(\, K^\phi\partial_\mu\phi^\dagger
 -\frac{i}{2}\,K_\phi^\phi \psi\,\sigma^\mu\,\opsi\,\big)
\Big\}\nonumber\\
&+&
F\,\frac{\partial \phi_X}{\partial\phi}
-\frac{1}{2}\,\psi\psi\,\frac{\partial^2\phi_X}{\partial \phi^2}
=
F\,\frac{\partial \phi_X}{\partial\phi}
-\frac{1}{2}\,\psi\psi\,\frac{\partial^2\phi_X}{\partial \phi^2}
+\cO(\partial_\mu)
\label{case1}
\eea

\medskip\noindent
where $\cO(\partial_\mu)$ terms vanish in the infrared limit.

Using the eqs of motion one  integrates out the sgoldstino $\phi$,
 to find the expression of $X$ in the low energy limit of decoupling
this field. Before doing so, let us find its mass.
The scalar potential is
\bea\label{min}
V=W_\phi\,W^\phi\,(K^{-1})^\phi_\phi
\eea

\medskip\noindent
The minimum conditions give
$k_{\phi\phi}^\phi\,w_\phi = w_{\phi\phi}\,k_\phi^\phi$ and
$k^{\phi\phi}_\phi\,w^\phi = w^{\phi\phi}\,k^\phi_\phi$ 
where $k_\phi^\phi, w_\phi$, etc denote the values of 
$K_\phi^\phi, W_\phi$, etc evaluated
on the ground state, i.e. $k_\phi$,  $w_{\phi\phi}$, $k_{\phi}^{\phi\phi}$...,
 are numbers.

As mentioned, we take $W=f\,\Phi$, then $w_{\phi\phi}=0$ so
it follows that on the ground state  $k_{\phi\phi}^\phi=k_\phi^{\phi\phi}=0$
and the goldstino is indeed massless. The conditions for local minimum 
are\footnote{For a general $W$, eqs.(\ref{ws1}), (\ref{ws2}) are
$
\big(
k_{\phi\phi}^{\phi\phi}\,\vert w_\phi\vert^2 
\!-k_\phi^\phi\,\vert w_{\phi\phi}\vert^2)^2
\!\geq\!
\vert w_\phi\vert^2 \vert
w_\phi k_{\phi\phi\phi}^\phi -\! k_\phi^\phi w_{\phi\phi\phi}
\vert^2$
and\newline
$m_{1,2}^2=
- (k_\phi^\phi)^{-2}\,
\big(\,
\vert w_\phi\vert^2\,k_{\phi\phi}^{\phi\phi}
-k_\phi^\phi\,\vert w_{\phi\phi}\vert^2
\pm 
\vert w_\phi \vert\,
\vert \, k_{\phi\phi\phi}^\phi\,w_\phi 
- k_\phi^\phi \,w_{\phi\phi\phi}\vert\,\big)
$.}
\medskip
\bea\label{ws1}
&&
\big(\,
k_{\phi\phi}^{\phi\phi}\,)^2-\vert\,k_{\phi\phi\phi}^\phi\vert^2\geq 0,\qquad
k_{\phi\phi}^{\phi\phi}<0
\eea

\medskip\noindent
and the scalar masses of real component fields of sgoldstino 
$\phi=1/\sqrt 2 (\varphi_1+i\varphi_2)$ are
\medskip
\bea\label{ws2}
m_{1,2}^2=
- (k_\phi^\phi)^{-2}\,f^2
\big(\,
k_{\phi\phi}^{\phi\phi} \pm \vert \, k_{\phi\phi\phi}^\phi \vert\,\big)
\eea

\medskip
Assuming that $K$ is such as $m_{1,2}^2$ are both positive and since $\psi$ is massless,
we can integrate out the sgoldstino via the equations of motion at 
zero momentum.  From our general Lagrangian for one field only
we find the eqs of motion\footnote{For $\phi$ and any $W$ this is
$K_{\phi\phi}^\phi F F^\dagger\!
-1/2 ( K_{\phi\phi\phi}^\phi \psi\psi F^\dagger
+ K_{\phi\phi}^{\phi\phi} \opsi\opsi F )
+ 1/4 K_{\phi\phi\phi}^{\phi\phi} \psi\psi \opsi\opsi
+W_{\phi\phi} F - 1/2 W_{\phi\phi\phi} \psi\psi\!=\!0
\label{generalphi}
$}
for $\phi, \phi^\dagger$ which we combine to obtain
\medskip
\be
K_{\phi\phi}^\phi\,K_\phi^{\phi\phi\phi}
- K_{\phi}^{\phi\phi}\,K_{\phi\phi}^{\phi\phi}
=\frac{\psi\psi}{2 F}\,
\Big[K_{\phi\phi\phi}^\phi \vert^2-(K_{\phi\phi}^{\phi\phi})^2\Big]
-
\frac{\psi\psi\,\opsi\opsi}{4\,F\,F^\dagger}
\,
\Big[K_{\phi\phi\phi}^{\phi\phi}\,K_\phi^{\phi\phi\phi}
- 
K_{\phi\phi}^{\phi\phi}
K_{\phi\phi}^{\phi\phi\phi}
\Big]
\ee

\medskip\noindent
One  expands this about the ground state $\langle\phi\rangle=0$, 
up to linear fluctuations in $\phi$, to find
\medskip
\bea\label{ro}
\phi=\frac{\psi\psi}{2\,F} -
\frac{\psi\psi\,\opsi\opsi}{4\,F\,F^\dagger}\,\,\,
\frac{k_{\phi\phi\phi}^{\phi\phi}\,k_\phi^{\phi\phi\phi}- 
k_{\phi\phi}^{\phi\phi}\,
k^{\phi\phi\phi}_{\phi\phi}\,
}{
\vert k_\phi^{\phi\phi\phi}\vert^2- (k_{\phi\phi}^{\phi\phi})^2}
+\cO(\phi^2;\phi^{\dagger 2};\phi\phi^\dagger)
\eea

\medskip\noindent 
According to (\ref{ws2}), the denominator in the middle term is proportional 
to the product of the two masses $- m^2_1 m^2_2/f^4$  and this presence
will be  encountered again in the case of more fields.
Relative to the denominator, the numerator of the same term has an extra $1/\Lambda$
due to extra derivative.
However, the error $\cO(\phi\phi^\dagger)$ simply makes the 
 coefficient of the middle term undetermined.
So the final result is, after eliminating $F$  (by $F\ra -f/k^\phi_\phi$):
\medskip
\bea
\phi=- (k^\phi_\phi)\,\frac{\psi\psi}{2 f}+\cO(1/\Lambda)
\eea

\medskip
We can use eq.(\ref{ro}) in eq.(\ref{ro2}) 
to compute $X$, after integrating out the sgoldstino $\phi$. 
The result is, for $W=f\,\Phi$ and  $K^\phi=\phi+\cO(1/\Lambda)$
\medskip
\bea\label{op1}
X=(4\,f+4/3\,F^\dagger)\,\Big[\frac{\psi\psi}{2\,F}+\sqrt 2\,\theta\psi 
+ F\,\theta\theta\Big]+\cO(1/\Lambda)
\eea

\medskip\noindent
which satisfies offshell $X^2=0$ up to terms $\cO(1/\Lambda)$ that
vanish in the IR limit. Onshell
\medskip
\bea\label{op2}
X=-(8/3\,f)\,\Big[\frac{\psi\psi}{2\,f}-\sqrt 2\,\theta\psi 
+ f\,\theta\theta\Big]+\cO(1/\Lambda)
\eea

\medskip\noindent
This concludes our review of the case of one superfield present only,
the goldstino itself.

\subsection{A simple example.}
\label{ae}

To illustrate the previous results, let us briefly 
apply them to some particular model. First we review the
model in \cite{SK}, with 
\medskip
\bea\label{st1}
K=\Phi^\dagger\Phi-\frac{c}{\Lambda^2}\,\Phi^2\,\Phi^{\dagger 2}
-\frac{\tilde c}{\Lambda^2}\,(\Phi^3\,\Phi^\dagger +\Phi\,\Phi^{\dagger 3}\,)
+\cO(1/\Lambda^3),
\quad W=f\,\Phi
\eea

\medskip\noindent
The higher dimensional D-terms ensure that the sgoldstino acquires a mass, 
to decouple in the IR, while the goldstino remains massless. Indeed,  from (\ref{min})
one finds a scalar potential 
\medskip
\bea
V=f^2\,\big[
1+4 \,c/\Lambda^2\,\,\phi^\dagger\phi+3/\Lambda^2\,\,(\tilde c\, \phi^2+h.c.)
+\cO(1/\Lambda^3)
\big]
\eea

\medskip\noindent
From this or from (\ref{ws2}) the masses are
$m_{1,2}^2=f^2\,(4\,c\pm 6 \tilde c)/\Lambda^2$ and with  the choice 
$\vert\tilde c\vert < (2/3)\,c$ one ensures stability and positive scalar (masses)$^2$.
Using the eqs of motion at zero momentum for $\phi$, one finds \cite{SK}
\bea\label{rr2}
\phi=\frac{\psi\psi}{2F}+\cO(1/\Lambda)
\eea

\medskip\noindent
in agreement with the previous general results.

\subsection{O'Raifeartaigh model with small supersymmetry breaking.}\label{or}

Let us make a side remark.  One may ask how to generate the higher dimensional 
terms in $K$ that ensure that the sgoldstino becomes massive, together 
with a linear superpotential which brings the Susy breaking, while 
the goldstino fermion remains massless. This can be done in a 
standard O'Raifeartaigh model. 
In such model, at tree level the sgoldstino is massless, but it 
acquires a mass via a
one loop renormalization of  $K$, induced by the other (massive)
superfields of the model. To see this, consider an O'Raifeartaigh model with
\bea\label{ae1}
K = \Phi_1^\dagger\Phi_1+ \Phi_2^\dagger\Phi_2+\Phi_3^\dagger\Phi_3,
\qquad
W = \frac{1}{2}\,h\,\Phi_1\,\Phi_2^2+ m_s\,\Phi_2\Phi_3 
+f\,\Phi_1
\eea

\medskip\noindent
so $\Phi_{2,3}$ have a large Susy mass ($m_{s}$) while $\Phi_1$ is massless.
One  computes the one-loop correction to the Kahler potential of $\Phi_1$ 
(this becomes the goldstino superfield) and then integrate 
out  the two massive superfields $\Phi_{2,3}$. The result is
shown below, under the simplifying 
assumption of small Susy breaking i.e. $f\,h < \vert m_s\vert^2$. For details
see Appendix A of \cite{Intriligator:2006dd}. One finds
\bea
K=\Phi_1^\dagger\Phi_1-\epsilon\,(\Phi_1^\dagger\Phi_1)^2+\cO(\epsilon^2),\qquad
W_{eff}=f\,\Phi_1,
\qquad
{\rm with}
\qquad 
\epsilon=\frac{1}{12}\Big(\frac{h^2}{4\,\pi}\Big)^2\,\frac{1}{\vert m_s \vert^2}
\eea

\medskip\noindent
For a reliable effective theory approach,  the mass of the sgoldstino
which is  $m_1^2=4\, \epsilon\,f^2$ should be of order $\sim f$,
which ultimately means $h^2\sim \cO(4\pi)$ i.e. a nearly strongly coupled regime. 
As seen from the previous example (\ref{st1}) with  
 $\tilde c=0$ and $c/\Lambda^2\rightarrow \epsilon$,
one finds that $ \phi^1={\psi^1\psi^1}/(2\,F^1)$, in agreement
with the general discussion.
 Other methods to generate a mass term for sgoldstino may be possible and
 in general strong dynamics is preferred.

\subsection{The results using the method of constrained superfields.}

The method of constrained superfields takes the property
$X^2=0$  that we saw is satisfied in IR after integrating out the 
sgoldstino $\phi$ and actually imposes it, from the beginning, as an input
constraint for the model. In fact it is enough to impose a weaker condition 
\bea\label{wqq}
F_{X^2}=0,\quad \Ra\quad \phi_X=\frac{\psi_X\psi_X}{2\,F_X}\quad \Ra\quad
\phi_X=\frac{\psi\psi}{2\,F}\,\frac{\partial\phi_X}{\partial \phi}
\eea

\medskip\noindent
So $F_{X^2}=0$ implies  $X^2=0$.
The last eq in (\ref{wqq}) is obtained after a Taylor 
expansion of the denominator in the previous step, 
in which we used $F_X$ of  (\ref{case1}).  One then finds
\medskip
\bea\label{rr}
W(\phi)+\frac{1}{3}\,\Big[ K^\phi\,F^\dagger
- \frac{1}{2} \, K^{\phi\phi}\,\,\opsi\opsi\Big]
=
\frac{\psi\psi}{2\,F}\,\Big[ W_\phi +\frac{1}{3}\,
\Big( K_\phi^\phi\,F^\dagger -\frac{1}{2}\, K_\phi^{\phi\phi}\,
\opsi\opsi\Big)\Big]
\eea

\medskip\noindent
which for $W=f\Phi$ gives 
\medskip
\bea\label{ooff}
\phi=-\frac{K^\phi\,F^\dagger}{3\,f}
+
\frac{K^{\phi\phi}}{6\,f}\,\,\opsi\opsi
+
\frac{1}{2\,F}\,\Big[ 1+\frac{1}{3\,f}\,K_\phi^\phi\,F^\dagger\Big]\,\psi\psi
-
\frac{K_\phi^{\phi\phi}}{12\,f\,F}\,\,\psi\psi\,\opsi\opsi
\eea

\medskip\noindent
Since all $K_\phi^\phi$, $K_{\phi}^{\phi\phi}$, $K^\phi$ depend on $\phi$, $\phi^\dagger$,
this is an implicit expression of $\phi$ in terms of the $\psi$, for any $K$
and can be used in applications. 
We take canonical kinetic terms $K=\Phi^\dagger\Phi+\cO(1/\Lambda)$ 
and  do not need to detail $\cO(1/\Lambda)$ terms, but only ensure 
that  they render $\phi$ massive as seen earlier.
From (\ref{ooff}) one finds
\bea
\label{off1}
\phi=\frac{\psi\psi}{2\,F}+\cO(1/\Lambda)
\eea
valid  offshell. 
Eq.(\ref{off1})  implies that $\Phi^2=\cO(1/\Lambda)$, 
which together with  $K=\Phi^\dagger\Phi+\cO(1/\Lambda)$
can reproduce in IR the Akulov-Volkov action  for $\Phi$.

Going to onshell supersymmetry, 
eq.(\ref{ooff})  gives 
\medskip
\bea\label{solution}
\phi=\frac{K^\phi}{3\,K_\phi^\phi}-
\frac{\opsi\opsi}{6\,f\,K_\phi^\phi}\,
\Big[
 K^\phi\,K_\phi^{\phi\phi}- K^{\phi\phi}\,K_\phi^\phi 
\Big]
-\frac{\psi\psi}{3\,f}\,K_\phi^\phi.
\eea

\medskip\noindent
This is the  implicit expression for the sgoldstino $\phi$ 
as a goldstino composite. It can be used  for specific applications, such as for example
\cite{AlvarezGaume:2010rt}.
There is dependence on both $\psi\psi$ and $\opsi\opsi$, 
but also possible dependence  on $\phi^\dagger$, so for specific 
$K$ one solves this for $\phi,\phi^\dagger$ in terms of fermion composites.
Taking $K=\Phi^\dagger\Phi+\cO(1/\Lambda)$ where $1/\Lambda$ suppresses the
dimensionful derivatives of $K$, 
from  (\ref{solution}) one recovers  (\ref{off1}) with 
$F$ replaced by $-f$.

As an application to Section~\ref{ae}, using (\ref{ooff}) 
in which we replace  the  derivatives of $K$ by their expressions derived 
with (\ref{st1}), we find an eq for $\phi$ (not shown), 
that is solved  easily by trying a solution  of the form
$\phi=a_1\,\psi\psi+a_2\,\opsi\opsi+a_3\,\psi\psi\opsi\opsi$ 
($a_{1,2,3}$ are constants to be identified). The final offshell solution 
turns out to be exactly eq.(\ref{rr2}) found instead via the eqs of motion.
Similar considerations apply for Section~\ref{or} whose result 
$\phi^1=\psi^1\psi^1/(2F^1)$ is also recovered from (\ref{ooff}).
This concludes our review of the one-field case.

\section{Goldstino and sgoldstino in the presence of matter fields.}
\label{twofields}

\subsection{General results.}

We proceed to study more realistic cases with a goldstino superfield in the
presence of matter superfields. The starting point is $L$ of (\ref{LKW}) 
in which sgoldstino and scalar matter fields acquire mass through
higher dimensional Kahler terms and decouple at low energy. It is our 
purpose to examine the expressions of these scalar fields
at the low energies, as functions (composites) of the light fermions present
(goldstino and matter fermions). 
Using these expressions, we then examine at low energy,
the expression  of  $X$ and its properties.

To this end, from  eq.(\ref{X})  we identify,
in the limit of vanishing space-time derivatives $\partial_\mu\ra 0$,   
the components of $X=(\phi_X,\psi_X,F_X)$:
\medskip
\bea
\label{compX}
\phi_X&=& 4\,W(\phi^i)+\frac{4}{3}\,\Big[K^j\,F_j^\dagger
 -\frac{1}{2} K^{ij}\,\opsi_i\opsi_j\Big]
\nonumber\\[4pt]
\psi_X&=& 
\psi^k\,\frac{\partial \phi_X}{\partial \phi^k}
-\frac{4\,i}{3}\sigma^\mu\,\big( K^j\,\partial_\mu\opsi_j
+K^{ij}\,\opsi_j\partial_\mu\phi_i^\dagger\big)
=
\psi^k\,\frac{\partial \phi_X}{\partial \phi^k}+\cO(\partial_\mu),
\nonumber\\[4pt]
F_X\! & = &\!\!\!
 F^i\,\frac{\partial \phi_X}{\partial \phi^i}\! -\!
\frac{1}{2} \psi^i\psi^j
\frac{\partial^2 \phi_X}{\partial \phi^i\partial\phi^j}
\!+\!\frac{4}{3}\,\Big\{
K_i^j\,\Big[\partial_\mu\phi^i\partial^\mu \phi_j^\dagger
+\frac{i}{2}\,\big(\psi^i\sigma^\mu D_\mu\opsi_j 
\!- \! D_\mu \psi^i\,\sigma^\mu \opsi_j\big)\Big]
\nonumber\\
&-&
\partial_\mu
\Big(
K^j\,\partial_\mu \phi_j^\dagger
-\frac{i}{2} K_i^j\psi^i\sigma^\mu\opsi_j\Big)
\Big\}
\,\, = \, F^i\,\frac{\partial \phi_X}{\partial \phi^i} -\frac{1}{2}\, \psi^i\psi^j\,
\frac{\partial^2 \phi_X}{\partial \phi^i\partial\phi^j}
+\cO(\partial_\mu)\qquad\qquad
\eea

\medskip\noindent
To evaluate $X$ after the decoupling of the sgoldstino/scalar matter fields, 
we consider, for simplicity, the Lagrangian (\ref{LKW}) for only two (super)fields, 
one goldstino and one matter scalar field, in order $\cO(1/\Lambda^3)$; we 
thus neglect corrections with more than four derivatives of  $K$. 
We also assume, for simplicity, that 
\bea
W=f\,\Phi_1,
\eea
 so $\Phi_1$ breaks supersymmetry.
The scalars $\phi^{1,2}$ acquire mass via higher dimensional D-terms and decouple
at low energy,  while with above $W$, $\psi^1$  remains massless 
(is the goldstino\footnote{See  next section for an example.}).

Let us integrate the scalars, by using the eqs of motion for
 $\phi^\dagger_{m}$,  $m=1,2$, which (at zero momentum) have the form
\medskip
\bea\label{sts}
\qquad\quad
K_i^{j1} F^i\,F_j^\dagger - \frac{1}{2}\,K_{ij}^{k1}\,\psi^i\psi^j
\,F_k^\dagger - \frac{1}{2}\,K_k^{ij1}\,\opsi_i\opsi_j\,F^k
+\cO(1/\Lambda^3)&=&0, 
\nonumber\\
K_i^{j2} F^i\,F_j^\dagger - \frac{1}{2}\,K_{ij}^{k2}\,\psi^i\psi^j
\,F_k^\dagger- \frac{1}{2}\,K_k^{ij2}\,\opsi_i\opsi_j\,F^k
 +\cO(1/\Lambda^3)&=&0, \,\,\,
i,j,k=1, 2.\quad
\eea

\medskip\noindent
To solve this for $\phi^{1,2}$ 
one expands (\ref{sts}) about the ground state (assumed $\langle\phi_i\rangle=0$),
so the field dependent Kahler  derivatives become
\medskip
\bea\label{ghh}
K_i^{jm}&=&k_i^{jm}+\phi^1\,k_{i1}^{jm}+\phi^2\,k_{i2}^{jm}+\cdots
\qquad\qquad
\nonumber\\[4pt]
K_{il}^{jm}&=&k_{il}^{jm}
+\cdots,
\qquad
K_{k}^{ijl}=k_k^{ijl}+\cdots
\eea

\medskip\noindent
where the dots represent contributions which are of
higher order $\cO(1/\Lambda^3)$.
As in the one-field case,
$k_{i\cdots}^{j\cdots}$ denote numerical values of the field-dependent 
Kahler derivatives $K^{i\cdots}_{j\cdots}$ on the ground state.
To simplify the calculation we work in normal coordinates basis,
where $k_i^{jm}=k_k^{ijl}=0$, $k_i^j=\delta_i^j$, $R_{ij}^{kl}=k_{ij}^{kl}$ 
for the curvature tensor, etc. Then the solution of system 
(\ref{sts}) for $\phi^{1,2}$ is, after using the permutation 
symmetries of some of the indices:
\medskip
\bea\label{eom}
\phi^1 \ &=&
\frac{\psi^1\psi^1}{2 F^1} \ - \ \frac{c_1}{2 F^1} (F^2 \psi^1 - F^1 \psi^2)^2
+\cO(1/\Lambda)\quad
\nonumber \\
\quad
\phi^2 \ &=& 
 \frac{\psi^2 \psi^2}{2 F^2} \ - \ \frac{c_2}{2 F^2} (F^2 \psi^1 - F^1 \psi^2)^2
+\cO(1/\Lambda)
\label{consk5}
\eea
where $c_{1,2}$ are
\bea\label{cof}
c_1 &=& \frac{\det \big[\, k_{2\,m}^{k\,n}\,F_k^\dagger\,\big]}{
\det\big[\, k_{i\,l}^{j\,p}\,F^i\,F_{j}^\dagger\,\big]}
\qquad
c_2 =
 \frac{\det \big[\, k_{1\,m}^{k\,n}\,F_k^\dagger\,\big]}{
\det\big[\, k_{i\,l}^{j\,p}\,F_i\,F_j^\dagger\,\big]}
\eea

\medskip\noindent
In $c_{1,2}$ the free indices $\{m,n\}$, $\{l,p\}$
are of the 2x2 matrices whose determinant is  evaluated.
The $\cO(1/\Lambda)$ correction originates from
eq.(\ref{sts}) with (\ref{ghh}). In $c_{1,2}$ one
can replace $k^{ij}_{lm}$ by corresponding curvature tensor $R^{ij}_{lm}$, 
to obtain the result for general coordinates\footnote{The relation
is
$R_{ij}^{kl}=k_{ij}^{kl}-k_{ij}^\rho\,(k^{-1})_\rho^n\,k_n^{kl}$; in 
normal coordinates $k_{ij}^k=0=k_k^{ij}$, so $R_{ij}^{kl}=k_{ij}^{kl}$. 
\newline
In  the complex geometry convention, $R_{ij}^{kl}$ is actually replaced by
$(R_{\overline k i})_{\overline l\,j}=K_{ij\,\overline l \overline k}
-K_{ij\overline\rho}\,K^{\overline\rho n}\,K_{n\overline k\overline l}$.
}. 
These results for $\phi^{1,2}$ together with corresponding $\psi^{1,2}$ and
$F^{1,2}$ define non-linear superfields $\Phi_{1,2}$  that can couple offshell 
the goldstino to matter, see for example Section~5 
of \cite{Dudas:2011kt}.

Intriguingly,  one notices by direct calculation that the scalar 
fields expressions in (\ref{eom}) are such as  that the corresponding
superfields of components $\Phi_i=(\phi^i,\psi^i,F^i)$ with $i=1,2$ and with 
$\phi^i$ as in (\ref{eom}) satisfy, for {\it arbitrary} $c_{1,2}$, 
the following generalized, higher order polynomial superfield constraints
\medskip
\bea\label{cc1p}
\Phi_1^3=\Phi_1^2\,\Phi_2=\Phi_1\,\Phi_2^2=\Phi_2^3=0
\eea

\medskip\noindent
which are valid offshell in IR (when we ignore $\cO(1/\Lambda)$ in the rhs of 
(\ref{eom})). To check (\ref{cc1p}) one shows by direct calculation
 that their scalar, 
fermion and auxiliary  components vanish, provided that $\phi^{1,2}$ have the 
form shown. This  property was noticed recently in
\cite{Dudas:2011kt} for particular cases, and as shown above, is actually
valid for general $K$.  Note also that offshell
\medskip
\bea\label{cc2p}
\Phi_1^2\not=0,\qquad \Phi_1\Phi_2\not =0
\eea

Let us also present some onshell results.
Using that the scalar potential is $V=(K^{-1})^i_j\,W_i\,W^j$ one can show that the
denominators of $c_{1,2}$ are, with $F^j$ replaced by their vev's, the
determinant of the squared mass matrix in the bosonic sector
of sgoldstino  ($\phi^1$) and scalar matter field\footnote{
The trace of the mass matrix is $-2\,R_{il}^{im}\,F_m^\dagger\,F^l$ \cite{GG}.}
($\phi^2$).
As shown by the scalar potential, these acquire mass via the Kahler metric. 
Going  to onshell supersymmetry, one finds
\medskip
\bea\label{conshell}
c_1=\frac{1}{f^2}\frac{\det \big[ R_{2m}^{1l}\big]}{\det\big[ R_{1m}^{1l}\big]},
\qquad c_2=\frac{1}{f^2}
\eea

\medskip\noindent
so the onshell result for $\phi^{1,2}$ is
\medskip
\bea
\phi^1&=&-\frac{\psi^1\psi^1}{2\,f}
+   \frac{\det \big[ R_{2m}^{1l}\big]}{\det\big[ R_{1m}^{1l}\big]}
\frac{\psi^2\psi^2}{2\,f}+\cO(1/\Lambda)
\nonumber\\[4pt]
\phi^2&=&-\frac{\psi^1\psi^2}{f}+\cO(1/\Lambda)
\label{osh}
\eea

\medskip\noindent
Note that the ratio of the two determinants is dimensionless and
 independent of the UV scale $\Lambda$. The absence in (\ref{osh}) of 
any similarity  in the structure of $\phi^{1,2}$  apparent in previous
(\ref{eom}) is due to  $c_{1,2}$ and to the fact 
that only $F^1$ is non-vanishing onshell in the approximation considered. 
In the presence of more matter fields, the solutions $\phi^j$, $j\geq 2$ 
have a similar form. Note that even onshell, $\Phi_1^2\not=0$.
In the {\it formal} limit of infinite scalar masses, onshell $c_1=0$.
With (\ref{eom}) one can investigate 
the properties of  $X$  after decoupling $\phi^{1,2}$ (Section~\ref{Xp}).

\subsection{A simple example.}\label{simpleex}

Let us first illustrate the results for $\phi^{1,2}$ of (\ref{eom})
for a particular model \cite{Dudas:2011kt}  with
\medskip
\bea
K
&=&
\Phi^\dagger_1\Phi_1
+\Phi_2^\dagger\Phi_2
-\epsilon_1 (\Phi_1^\dagger \Phi_1)^2
-\epsilon_2  (\Phi_2^\dagger\Phi_2)^2
\nonumber\\[6pt]
&& \qquad - \,\, \epsilon_3 (\Phi_1^\dagger\Phi_1) (\Phi_2^\dagger\Phi_2)
- \epsilon_4 \, \big[(\Phi_1^\dagger)^2 \Phi_2^2 + h.c.\big]
+ \cO(1/\Lambda^3)
\eea
and superpotential
\bea
W&=& f\,\Phi_1,\qquad\qquad \epsilon_i=\cO(1/\Lambda^2)\qquad\qquad
\eea

\medskip\noindent
The scalar potential is 
\medskip
\bea
V=(K^{-1})^k_\rho\,W_k\,W^\rho= (K^{-1})^1_1 f^2=f^2\,
\big(
1+\epsilon_3\, \vert\phi^2\vert^2+
4\,\epsilon_1\,\vert\phi^1\vert^2 \big)
\eea

\medskip\noindent
therefore $m_{\phi_1}^2=\epsilon_3\,f^2$, $m_{\phi_2}^2=4\,\epsilon_1\,f^2$ for the 
sgoldstino ($\phi^1$) and scalar matter field ($\phi^2$), respectively (we choose
$\epsilon_{1,3}>0$) and the ground state is indeed at $\phi_{1,2}=0$.

The solution for $\phi^{1,2}$ is that of (\ref{consk5}); using
eq.(\ref{cof}) one finds \cite{Dudas:2011kt}
\medskip
\begin{eqnarray}
 c_1 \ &=& \frac{1}{\Delta}\, \epsilon_3\, (\epsilon_2 F^{\dagger 2}_2 -
\epsilon_4 F^{\dagger 2}_1)=  
-\frac{\epsilon_4}{\epsilon_1}\frac{1}{ F_1^2}+\cO(1/\vert F_1\vert^4)
\newline\nonumber\\
c_2 \ &=& \frac{1}{\Delta}\,\epsilon_3\, (\epsilon_1 F^{\dagger 2}_1
 - \epsilon_4 F^{\dagger 2}_2) 
=\frac{1}{F_1^2}+\cO(1/\vert F_1\vert^4)
\ , \quad {\rm where } \nonumber \\
  \Delta  &= & 
 \ \epsilon_3 F_1^2  (\epsilon_1 F^{\dagger 2}_1 
- \epsilon_4 F^{\dagger 2}_2)
+ \epsilon_3 F_2^2  (\epsilon_2 F^{\dagger 2}_2 - \epsilon_4 F^{\dagger 2}_1)
+ 4 (\epsilon_1 \epsilon_2 - \epsilon_4^2) |F_1|^2 |F_2|^2 
\label{consk8}
\end{eqnarray}

\medskip\noindent
The offshell result for $\phi^{1,2}$ is given in  (\ref{consk5}) with these 
$c_{1,2}$ and has
$\cO(1/\Lambda)$ correction that originates from the eqs 
of motion for $\phi^{1,2}$ that
are valid in $(1/\epsilon_1)\times \cO(1/\Lambda^3)=\cO(1/\Lambda)$ (since 
$\epsilon_1$ is multiplying  $\phi^{1,2}$ in these eqs).
The rhs expansion of $c_{1,2}$ in $1/\vert F_1\vert$ is allowed onshell, after
taking account that onshell $F^1=-f+\cO(\epsilon_i)$ and $F^2=\cO(\epsilon_i)$. Note that
$c_{1,2}$ are ratios of $\epsilon_i$, so $c_{1,2}=\cO(1/\Lambda^0)$. 
The onshell result is then
\medskip
\bea\label{onshell19}
\phi^1&=&
-\frac{\psi^1\psi^1}{2\,f}
-\frac{\epsilon_4}{\epsilon_1}\,\frac{\psi^2\psi^2}{2\,f} +\cO(1/\Lambda)
\nonumber\\
\phi^2&=&
-\frac{\psi^1\psi^2}{f}
+\cO(1/\Lambda)
\eea

\medskip\noindent
in agreement with (\ref{osh}). Obviously, $\Phi_1^2\not=0$ since $(\phi^1)^2\not=0$.

\subsection{The properties of the field $X$ after integrating out the scalar fields.}
\label{Xp}

Returning now to the field $X\!=\!(\phi_X,\psi_X,F_X)$  of (\ref{compX}),
we use the solution of the eqs of motion for $\phi^{1,2}$ of (\ref{eom}),
 to find the expression of $X$ after integrating out these scalars.
For canonical kinetic terms $K=\Phi_1^\dagger\Phi_1+
\Phi_2^\dagger\Phi_2+\cO(1/\Lambda)$ and with $W=f\Phi_1$ we find after some 
algebra
\medskip
\bea\label{Xsquared}
X^2&=&
\rho\,\Big[ 
(\psi^1\psi^1)\,(\psi^2\psi^2)
-
4\,\sqrt 2  (\psi^1\psi^2)\,(F^1\, \theta\psi^2+F^2\,\theta\psi^1)
+
2\,(\theta\theta)\,(\psi^1\,F^2-\psi^2\,F^1)^2
\Big]\nonumber\\[6pt]
&+&\cO(1/\Lambda)
\eea

\medskip\noindent
where $\cO(1/\Lambda)$ involves terms that contain $\overline\psi$;
from (\ref{Xsquared}) one can read the
components of $X^2=(\phi_{X^2}, \psi_{X^2},F_{X^2})$. Also

\medskip\noindent
\bea\label{rho}
\rho& \equiv & \frac{1}{2 F^1\,F^2}\,\Big[\sigma_1\sigma_2-
(F^1\,\sigma_1+F^2\,\sigma_2)\,(c_1\sigma_1\,F^2+c_2\,\sigma_2\,F^1)\Big]
+\cO(1/\Lambda)
\nonumber\\[6pt]
\sigma_1&=& 4 f+4/3\,F_1^\dagger,\,\,\,\sigma_2=4/3\,F_2^\dagger
\eea

\medskip\noindent
where again $\cO(1/\Lambda)$ suppresses $\overline\psi$-dependent terms
and $c_{1,2}$ are those of (\ref{cof}).
Obviously $X^2\not= 0$, unless $\rho=0$. However, with $c_{1,2}$ of (\ref{cof})
this is not possible in general. In specific cases with particular
values for the individual Kahler curvature terms, one may have a 
vanishing $X^2$ but this
is not true in general. Thus the conjecture $X^2=0$ is not verified
in general.

Let us evaluate $X^2$ onshell. One finds, using (\ref{conshell}) that
\medskip
\bea\label{X2os}
\rho\,\Big\vert_{\rm onshell}=-\frac{1}{2}\,c_1\,(8/3\,f)^2
=
-\frac{32}{9}\,\frac{\det R_{2m}^{1l}}{\det R_{1m}^{1l}}+\cO(1/\Lambda)
\eea

\medskip\noindent
As a result
\medskip
\bea
X^2\Big\vert_{\rm onshell}=
-\frac{64}{9}\,\bigg[\frac{\det R_{2m}^{1l}}{\det R_{1m}^{1l}}\bigg]\,f\,
\,(\psi^2\psi^2)\,
\bigg[\,\frac{\psi^1\psi^1}{2\,f}
-\sqrt 2 \,(\theta\psi^1)
+
(\theta\theta)\,f\,\bigg]+\cO(1/\Lambda)
\eea

\medskip\noindent
which clearly does not vanish, except in specific cases when 
$c_1=0$, i.e. if the determinant in the numerator vanishes or
that in the denominator is infinite. The last factor in the rhs is the (onshell)
goldstino superfield in the absence of matter superfields and after integrating out the
sgoldstino, see (\ref{op2}).

So far we found that $X^2$ is not vanishing offshell or onshell. 
Next, let us investigate the offshell value of 
$X^3$. One finds using (\ref{compX})  together with
canonical kinetic terms for $\Phi_{1,2}$ and $W=f \Phi_1$, that 
\medskip
\bea
\phi_{X^3}&=&\phi_X\,\phi_{X^2}
\propto (\phi^1\,\sigma_1+\phi^2\,\sigma_2)\,(\psi^1\psi^1)\,(\psi^2\psi^2)
+\cO(1/\Lambda)=\cO(1/\Lambda)
\nonumber\\[6pt]
\psi_{X^3}&=&
3\,\phi_X^2\,\psi_X
=
(3/2)\,\phi_X\,\psi_{X^2}\propto
(3/2)\,(\phi^1\sigma_1+\phi^2\sigma_2)(\psi^1\psi^2)
(F^1\,\theta\psi^2+F^2\,\theta\psi^1)
=
\cO(1/\Lambda)
\nonumber\\[6pt]
F_{X^3}&=&
3\,\phi_X\,(F_X\,\phi_X-\psi_X\psi_X)
=
3 \,(\phi^1\sigma_1+\phi^2\sigma_2)\,\big[
(\phi^1\sigma_1+\phi^2\sigma_2)\,(F^1\sigma_1+F^2\sigma_2)
\nonumber\\[6pt]
&-& (\psi^1\sigma_1+\psi^2\sigma_2)^2\big]=\cO(1/\Lambda)\,\,
\eea

\medskip\noindent
The last step in the rhs of each of these eqs uses that $\phi^{1,2}$ contain
bilinears in $\psi^{1,2}$ as seen from (\ref{eom})
and also the expressions of the components of $X^2=(\phi_{X^2},\psi_{X^2},F_{X^2})$ as 
shown in (\ref{Xsquared}). Finally, for $F_{X^3}$ we used (\ref{eom}) 
(without replacing $c_{1,2}$ by their values (\ref{cof})). In conclusion, $X^3=0$ in 
the infrared (offshell-Susy).

One could reverse the arguments and take the property $X^3=0$ 
and consider it as an input condition, in a constrained superfields formalism, 
to identify
the goldstino superfield (and to replace the conjectured $X^2=0$ constraint, 
see last section). We chose not to do so, for the following reasons.
Let us remind that these results are for a vanishing ``improvement'' term in (\ref{eq9}).
 In the presence
of an arbitrary non-vanishing such term, even the property $X^3=0$ shown above
for 2 fields can be violated, since (\ref{compX}) is changed.
This stresses that the properties of $X$ are not uniquely 
defined, not even in onshell-Susy for $X$ or its powers. As a result, conclusions
derived from assuming some constraints on $X$ have to be regarded with due care.
Note that while the properties of $X$ such as $X^3=0$  are affected by
the improvement term, those of $\Phi_{1,2}$ superfields (\ref{cc2p}) 
are independent of this, and can be considered as more fundamental
and of more use in practice.
Finally, in the presence 
of more matter  fields ($n$) or (non-linear) superpotential terms, 
this condition is likely to change into $X^{n+1}=0$, see also
\cite{Dudas:2011kt}.

For future reference, we provide below onshell-supersymmetry results, without 
an approximation in $1/\Lambda$  and also for an arbitrary $W$, without 
integrating out the scalar fields $\phi^j$.  These are  obtained from 
the onshell structure of $\phi_X, F_X, \psi_X$. Using (\ref{compX}) one finds
\medskip
\bea\label{ons1}
\phi_X&=& \sigma+\sigma^{mn}\,\,\opsi_m\opsi_n,
\quad\qquad
\psi_X = \frac{8}{3}\,\psi^k\,W_k
\nonumber\\
F_X&=& \beta+\beta_{mn}\,\psi^m\psi^n+
\beta_{mn}^{kl}\,\,(\psi^m\psi^n)\,\,(\opsi_k\opsi_l)
\eea
where
\bea\label{ons2}
\sigma&=&4\,W-(4/3)\,\, K^l\,(K^{-1})^k_l\,\,W_k, 
\qquad
\sigma^{mn}=(2/3)\,\big[K^l\,(K^{-1})^k_l\,K_k^{mn}-K^{mn}\big]
\nonumber\\
\beta&=&-(8/3)\,\,W_m\,(K^{-1})^m_k\,\,W^k,\,\,
\quad
\beta_{mn}= 2\,\big[\,W_{k}\,\Gamma^k_{mn}-W_{mn}\big],\,\,
\quad
\beta_{mn}^{kl}=(1/3)\,\,R_{mn}^{kl}
\eea

\medskip\noindent
 $R_{mn}^{kl}$ denotes the curvature tensor,
$R_{mn}^{kl}=K_{mn}^{kl}-K_{mn}^j\, (K^{-1})_j^i\,K_i^{kl}$
and we considered an arbitrary superpotential $W$.
 Here all coefficients of the fermions are  scalar field-dependent.

\subsection{Another example.}\label{anotherex}

For more insight into the properties of the
field $X$ and the relation to  $\Phi_{1,2}$ consider
a particular model \cite{Dudas:2011kt} with
\bea
&& K \ =
\Phi_1^{\dagger} \Phi_1+\Phi_2^{\dagger} \Phi_2
 \ - \  \epsilon\ \big(
 \Phi_1^{\dagger} \Phi_1+\Phi_2^{\dagger} \Phi_2
\big)^2 \ , \qquad
W \ = \ f \ \Phi_1 \ ,
\label{goldstino1}
\eea
 One can easily extend this to $N$ fields
by extending in $K$ the two sums from $2$ to $N$ and 
also $f\Phi_1 \ra \sum_{i=1}^N f_i\Phi_i$ (similar for the formulae below).
Using a zero-momentum integration of the heavy scalars $\phi^{1,2}$,
one finds
\medskip
\bea
\Phi_i \ = \Big\{\,
\frac{1}{|F|^2} \ \Big(\psi^i - F^i \ \frac{{F_k^\dagger}\, \psi^k }{2 |F|^2} \Big)
\ {F_n^\dagger}\, \psi^n\, \Big\} \ + \ \sqrt{2} \ \theta \ \psi^i
\ + \ \theta^2 \ F^i \ , \quad
|F|^2=| F_1|^2+|F_2|^2.\label{goldstino2}
\eea

\medskip\noindent
where $\phi^i$ defined by the curly bracket is obtained from
eq.(\ref{eom}), (\ref{cof}).

As also seen in (\ref{cc2p}), note that
 $\Phi_i^2 \not=0$, $i=1,2$.  However, {\it define}
 the superfield
\bea
\tilde X \ = \ \frac{1}{|F|} \ {F^\dagger_i} \,\Phi^j \
 = \ \frac{({F_j^\dagger}\, \psi^j)^2}{2 |F|^3} \ + \ \sqrt{2} \ \theta
\ \frac{{F_j^\dagger}\, \psi^j}{|F|} \ + \ \theta^2 \ |F|
\eea

\medskip\noindent
Interestingly, this superfield satisfies the constraint $\tilde X^2=0$, with goldstino defined by 
the $\theta$ component. Further, any linear combination
$\Phi_i^\prime = c_{ij} \Phi_j$ satisfies the
constraint $\tilde X \Phi_i^\prime=0$. 
It should be recognized that this is a very symmetric example
and the constraint $\tilde X^2=0$ seen here is not  generally valid, 
as already known from previous example, Section~\ref{simpleex}.

The natural question is: what is the link of $\tilde X$ with $X$ of (\ref{eq9}),
 (\ref{compX}), or why $X^2\not=0$ in this case. 
Indeed, according to  (\ref{Xsquared}), (\ref{rho}) for this model  $X^2\not=0$,  
unless one formally takes $f=0$ (unacceptable) or, alternatively,
chooses a non-zero  improvement term that cancels $W(\phi_i)$ in $\phi_X$ of
(\ref{compX}) but leaves $\psi_X, F_X$ unchanged, up to space-time derivatives. 
Such an improvement term would require a non-local $Y$, and
is therefore not acceptable, see discussion and footnote 
immediately after eq.(\ref{eq9}). 
In conclusion $X$ of (\ref{eq9}), (\ref{compX})
 cannot satisfy $X^2=0$, even though
there does exist another, unrelated superfield $\tilde X$ with this property.
Nevertheless, the existence of $\tilde X$ with such property is
 specific to this
very special model.

\section{The constrained superfields method and its validity.}
\label{limits}

For a better understanding of the results so far, one would 
like to have a closer look at the infrared conjecture $X^2=0$ 
in the presence of matter fields \cite{SK}, as a definition for the 
sgoldstino.
For the sake of the argument, in this section we therefore assume  that this is 
true and then analyze its implications, to see more closely its validity limits,
by comparing with the results of the previous sections.
In fact it is sufficient to impose $F_{X^2}=0$.  One has
\medskip
\bea\label{w}
F_{X^2}=0\,\qquad \Ra\qquad 
\phi_X= \frac{\psi_X\psi_X}{2\,F_X}\qquad\Ra\qquad  X^2=0.
\eea

\medskip\noindent
After using (\ref{compX}), one obtains from the middle relation
\medskip
\bea\label{fix}
\phi_X=
\Big(\frac{\partial\phi_X}{\partial \phi^j} \,\psi^j\Big)^2\,
\,
\Big(2\,F^k\, \frac{\partial \phi_X}{\partial \phi^k}\Big)^{-1}
\,
\Big\{
1+\sum_{j\geq 1} \Big[\psi^k\psi^l\,
\frac{\partial^2\phi_X}{\partial \phi^k\partial\phi^l}
\,\,
\Big(2\,F^k\, \frac{\partial \phi_X}{\partial \phi^k}\Big)^{-1}
\Big]^j
\Big\}
\eea

\medskip\noindent
where $\cO(\partial_\mu)$ terms, that vanish in the IR,
are not displayed from now on. The sum over $j$ has actually a 
finite number of terms, for a large enough power of the Grassmann variables $\psi^k$.
Together with the definition of $\phi_X$ in (\ref{compX}) used to
 evaluate the derivatives,
(\ref{fix})  is an implicit  definition of  $\phi^1$ to be identified 
later with the sgoldstino ($\psi^1$ is the goldstino) if $W=f\Phi_1$.

To integrate out the scalars fields $\phi^j$ ($j\not=1$) other than sgoldstino, one 
imposes \cite{SK}
\medskip
\bea\label{ww}
F_{X\,\Phi_j}=0,\qquad \Ra\qquad \phi^j= \frac{\psi^j\,\psi_X}{F_X} -\phi_X\,\frac{F^j}{F_X}
\qquad \Ra\qquad X\,\Phi_j=0
\eea

\medskip\noindent
where $\Phi_j$ is an arbitrary matter superfield and eq.(\ref{w}) was used.
The middle relation can be re-written as
\medskip
\bea\label{swd}
\phi^j=\frac{\psi^j\,\psi^j}{2\,F^j}
-\frac{1}{2\,F^j}\,\frac{1}{F_X^2}\,(F^j\psi_X-F_X\psi^j)^2,
\qquad {\rm j:\,fixed,\,j\not=1.}
\eea

\medskip\noindent
This form  anticipates the structure of the offshell solution for 
scalar matter fields and sgoldstino (see later).
From this one finds
\medskip
\bea\label{fij}
\phi^j=\frac{\psi^j\psi^j}{2\,F^j}
-\frac{1}{2\,F^j}\,\frac{1}{F_X^2}\,
\Big[
(F^k\,\psi^j- F^j\,\psi^k)\,\frac{\partial\phi_X}{\partial \phi^k}
-
\frac{1}{2}\,(\psi^m\psi^k)\,\psi^j\,\,
\frac{\partial^2\phi_X}{\partial\phi^m\partial\phi^k}
\Big]^2
\eea

\medskip\noindent
where for the derivatives of $\phi_X$ one uses the first line in (\ref{compX}). 
Such derivatives can only bring 4-fermion or more corrections 
that are $\cO(1/\Lambda)$. Similar considerations for $\phi_X$ of (\ref{fix}).

Eq.(\ref{fix}), (\ref{fij}) are general results. They simplify 
if we ignore the $\cO(1/\Lambda)$ terms to give
\medskip
\bea\label{fr1}
\!\!\phi_X & = &\!\frac{\alpha_i\,\alpha_j}{2 \,F^k\,\alpha_k}\,\,
\psi^i\psi^j\,\,\Big\{ 1+\sum_{p\geq 1}  
\Big(\frac{2 \,W_{mn}}{F^l\,\alpha_l}\,\psi^m\psi^n\Big)^p
\Big\}+\cO(1/\Lambda),
\\
\!\phi^j &=&\!\!
\frac{\psi^j\psi^j}{2\,F^j}-
\frac{1}{2 F^j}
\big(F^k\alpha_k -2 W_{mn}\psi^m\psi^n\big)^{-2}
\big[ \alpha_k (F^k\psi^j-F^j\psi^k)
-2 W_{ls} (\psi^l\psi^s) \psi^j\big]^2
\!\!+\!\cO(1/\Lambda)\nonumber
\eea
where
\bea
\alpha_k = 4\,W_k+(4/3)\,K_k^m\,F_m^\dagger
\eea

\medskip\noindent
where $\cO(1/\Lambda)$ accounts for $\geq 4$ fermions terms.
In the first bracket of $\phi^j$ one Taylor expands to a finite series in spinors. 
If $W_{mn}\not=0$ four-fermion terms can be present even in order $\cO(1/\Lambda^0)$.
For $W_{mn}=0$, eq.(\ref{swd}) becomes
\medskip
\bea\label{opop}
\phi^j=
\frac{\psi^j\psi^j}{2\,F^j}-
\frac{1}{2\,F^j}
\frac{\big[\alpha_k\,(F^k\psi^j-F^j\psi^k)\big]^2}{
\big[F^k\alpha_k\big]^{2}}+\cO(1/\Lambda)
\eea

\medskip\noindent
which is a form that we shall use shortly.

\subsection{The case of two scalar fields $\phi^{1,2}$ - offshell results.}
\label{abs}

Let us now consider the implications of the above results
for the case of two fields only (goldstino plus a matter field) 
but for an arbitrary Kahler potential with canonical kinetic terms
and a ``standard'' goldstino superpotential: 
\bea\label{WK1}
K=\Phi_1^\dagger\Phi_1+\Phi_2^\dagger\Phi_2+\cO(1/\Lambda),
\qquad
W=f\,\Phi_1
\eea

\medskip\noindent
$\cO(1/\Lambda)$ terms stand for Kahler terms that
are higher dimensional and that we do not  need to specify explicitly here; 
they  involve derivatives $K_{ij}^m$, $K_{ij}^{kl}$, etc in  (\ref{LKW}), and are 
thus suppressed by powers of $1/\Lambda$. Such terms also give mass to 
sgoldstino $\phi^1$ and scalar matter field $\phi^2$ as we saw earlier. 
After using eqs.(\ref{fr1}), (\ref{opop}) with $W$ of  (\ref{WK1}), 
one finds
\medskip
\bea\label{off2}
\phi^1 & = & \frac{\psi^1\psi^1}{2\,F^1}
-
\frac{c_1}{2\, F^1}\,(F^2\,\psi^1-F^1\,\psi^2)^2+\cO(1/\Lambda)
\nonumber\\[6pt]
\phi^2\!\! & = & 
\frac{\psi^2\psi^2}{2\,F^2}
-
\frac{c_2}{2\, F^2}\,(F^2\,\psi^1-F^1\,\psi^2)^2
+\cO(1/\Lambda)
\eea

\medskip\noindent
where the ``new'' $c_{1,2}$ denote some functions of $F_{1,2}$:
\medskip
\bea\label{c12}
c_1  =\frac{\alpha_2^2}{(F^k\alpha_k)^2}=
\frac{[(1/3) F_2^\dagger]^2}{[F^k\,(\delta_{k1}\,f+1/3\,F_k^\dagger)]^2},
\qquad
c_2  =\frac{\alpha_1^2}{(F^k\alpha_k)^2}=
\frac{(f+1/3 F_1^\dagger)^2}{[F^k\,(\delta_{k1}\,f+1/3\,F_k^\dagger)]^2}
\eea

\medskip\noindent
This\footnote{A consistency check: with these values of $c_{1,2}$, 
$X^2\propto \rho$ of (\ref{X2os}) is indeed vanishing.}
 is the offshell  form of sgoldstino $\phi^1$ and  
scalar matter field $\phi^2$. Note the similarity with (\ref{swd}), (\ref{opop})
and the symmetry of $c_{1,2}$ in indices $1,2$, for the formal 
limit of vanishing supersymmetry scale $f$. 
The result in (\ref{off2}) takes into account that offshell and in the IR limit
$\phi_X$ is not $\phi_1$ but rather a combination of $\phi^1$ and $\phi^2$, 
with no relative suppression in $\Lambda$.
This is seen from the definition of $\phi_X$ which contains terms such as 
$\phi_X\supset K^kF_j^\dagger=\phi^1 \,F_1^\dagger+\phi^2\,F_2^\dagger+\cO(1/\Lambda)$ 
which involves both $\phi^{1,2}$.

Eqs.(\ref{off2}), (\ref{c12}) represent the main result of this section 
and should be compared to 
those in (\ref{eom}) with (\ref{cof}). One immediately sees that while
the structure of the solution is similar, the offshell values of
the coefficients $c_{1,2}$ are different in the two cases. 
The ultimate reason of this difference is due to the fact that
the conjecture $X^2=0$ in IR has a  limited validity
and is at the origin of this discrepancy.

As seen in the previous section, one notices that the scalar 
fields expressions in (\ref{off2}) are such  that the corresponding
superfields of components $\Phi_i=(\phi^i,\psi^i,F^i)$ with $i=1,2$ and with 
$\phi^i$ as in (\ref{off2}) satisfy, for {\it arbitrary} $c_{1,2}$, 
the higher order polynomial superfield constraints, valid offshell,
shown in (\ref{cc1p}), which are independent of the exact values of $c_{1,2}$
(and ignoring $\cO(1/\Lambda)$ in the rhs of (\ref{off2})).
Also notice that offshell (\ref{cc2p}) is also respected.

To understand the relation with the result in \cite{SK}
consider  the formal  limit of large scale of supersymmetry breaking ($\sqrt f$).
One  should be aware of the restricted validity of this limit, 
since we are already
within a $\cO(1/\Lambda)$ expansion, which can contain in particular
$\cO(\sqrt f/\Lambda)$ terms. 
Nevertheless, in this case (\ref{off2}) with (\ref{c12}) gives
\medskip
\bea
\phi^1 & = & \Big\{\frac{\psi^1\psi^1}{2\,F^1}
-\frac{1}{18\,f^2}\frac{F_2^{\dagger 2}}{(F^1)^3}\,(F^2\,\psi^1-F^1\,\psi^2)^2+
\cO(1/f^3)\Big\}+\cO(1/\Lambda)
\nonumber\\
\phi^2 & = & \Big\{\frac{\psi^1\psi^2}{F^1}-\frac{F^2\,\psi^1\psi^1}{2\,(F^1)^2}
+
\frac{1}{3\,f}\frac{F_2^\dagger}{(F^1)^3}\,(F^2\,\psi^1-F^1\,\psi^2)^2
\nonumber\\
&-&
\frac{1}{18\,f^2}\frac{F_2^\dagger}{(F^1)^4}\,
(2\,\vert F^1\vert^2+3\,\vert F^2\vert^2)
\,(F^2\,\psi^1-F^1\,\psi^2)^2+
\cO(1/f^3)\Big\}+\cO(1/\Lambda)
\eea

\medskip\noindent
{For infinite $f$ (i.e. order $1/f^0$)
only the first (first two) term in $\phi^1$ ($\phi^2$) 
contribute respectively, and one recovers the 
result  \cite{SK}
\bea
\phi^1=\psi^1\psi^1/(2F^1)\,\qquad
\phi^2=\psi^1\psi^2/F^1-F^2\,\psi^1\psi^1/(2 (F^1)^2)
\eea

Let us  now discuss some onshell-Susy results. Eq.(\ref{off2}) gives onshell
\medskip
\bea\label{onshell2}
\phi^1 = - {\psi^1\psi^1}/(2 \,f)+\cO(1/\Lambda),\qquad
\phi^2 = - {\psi^1\psi^2}/f+\cO(1/\Lambda)
\eea

\medskip\noindent
Note that in the onshell case $\phi_X= 4 f\phi_1- 4/3 K^j (K^{-1})^1_j\, f =
8/3 \phi_1\,f +\cO(1/\Lambda)$ so $\phi_X\propto \phi^1$
while offshell $\phi_X$ is a mixture of both  $\phi^{1,2}$ as already mentioned.
Note also that onshell  $\Phi_{1,2}$  
(i.e. auxiliary $F^{1,2}$ replaced by their solution of eqs of motion
and the scalar components as in (\ref{onshell2})) satisfy
\medskip
\bea
\Phi_1^2=\Phi_1\,\Phi_2=0 \,\,\,\textrm{in IR}
\eea

\medskip\noindent
 in contradiction with $\Phi_1^2\not=0$ of~(\ref{osh}).

The onshell results of the last two eqs can also be obtained using directly the 
onshell form of $X$, after imposing  $\phi_X=\psi_X\psi_X/(2\,F_X)$ that 
comes from the IR conjecture $X^2=0$. From this together with (\ref{ons1}), 
(\ref{ons2}) one finds
\medskip
\bea\label{ons3}
\phi_X=\frac{32}{9}\frac{\psi^k\psi^l\,\,W_k\,W_l}{
\beta+\beta_{mn}\,\,\psi^m\psi^n+\beta_{mn}^{kl}\,\,(\psi^m\psi^n)\,(\opsi_k\opsi_l)}
=
\frac{32}{9\,\beta}\,\psi^k\psi^l\,W_k\,W_l
+\cdots
\eea

\medskip\noindent
which can be Taylor expanded into a finite series.
For our simple case with only 2 fields (goldstino and a matter field)
with canonical kinetic terms in $K$ and  $W=f\,\Phi_1$ 
one finds
\medskip
\bea
\phi_X=\frac{32}{9}\frac{f^2}{\beta}\,\,\psi^1\psi^1+(\geq 4\,\,{\rm fermions})
\eea

\medskip\noindent
which has no $\psi^2\psi^2$ contribution in the onshell-Susy case. Using 
(\ref{ons1}), (\ref{ons2}) one finds
\medskip
\bea\label{ons4}
\phi^2 &=& \frac{8\,f}{3\,\beta}\,\,(\psi^1\psi^2) +\frac{32}{9}\,\frac{f^3}{\beta^2}\,
(K^{-1})^2_1\,\,(\psi^1\psi^1)+(\geq 4\,\,{\rm fermions})=
-\frac{\psi^1\psi^2}{f}+\cO(1/\Lambda)
\eea

\medskip\noindent
and that
\bea\label{ons5}
\phi_1=-\frac{\psi^1\psi^1}{f}+\cO(1/\Lambda)
\eea

\medskip\noindent
in agreement with (\ref{onshell2}). 
From this we find again that $\phi^1$ cannot contain, onshell,
a $\psi^2\psi^2$ fermionic pair, as also seen in  (\ref{onshell2}).
 This is in contradiction with the result found in 
 (\ref{conshell}), (\ref{osh}),  (\ref{onshell19}).
As a result the IR constraint $X^2=0$ 
leads to results that onshell/offshell are not correct, except when  $c_1=0$
when agreement with the results of Section~\ref{twofields} exists (onshell).

One possibility for $c_1=0$ is in the limit the denominator in $c_1$ that
is  proportional (onshell) to the scalar masses product (see (\ref{conshell})),
 is infinite, i.e. 
 the scalar masses that are projected out by  the constraints 
 are infinite. This may not be too  surprising, given that the 
constraints $X^2=0$ and $X\Phi_2=0$, not involving the curvature tensor, are
 not sensitive to the spectrum of integrated scalars, which must thus be decoupled. 
 
 A second possibility for $c_1=0$ is when  $\det (R^{1k}_{2l})=0$, 
in which case agreement with  Section~\ref{twofields}
is again obtained onshell, even when the masses of the integrated scalars are finite. 
 Such agreement exists  if the Kahler potential has a symmetry that 
 enforces $c_1\propto \det (R^{1k}_{2l})=0$, with $\det (R^{1k}_{1l})\not=0$ and finite. 
 For example one can consider a discrete R-symmetry, with $\Phi_{1,2}$ of different 
 R-charges $q_{1,2}$; if such symmetry of $L$ exists, 
 then these  conditions can indeed be realized, with $q_1\not=q_2$.
In particular,  different discrete R-charges for $\Phi_{1,2}$ forbid the term in 
$K\supset \epsilon_4 (\Phi_1^\dagger)^2\,(\Phi_2)^2+h.c.$ that is
present in (\ref{onshell19}) and then agreement with
Section~\ref{twofields} exists. Thus in the presence of such symmetry
the constrained superfields method based on the infrared constraint
$X^2=0$ can give correct results onshell. One should keep however in mind that
it is the {\it offshell} constraint that is relevant when coupling the goldstino
field to matter, in a superfield language.

\section{Conclusions.}\label{conc}

For a general nonlinear sigma model,
we studied the properties of the superfield $X$ that violates
the conservation of the Ferrara-Zumino  supercurrent and breaks supersymmetry 
and the conformal symmetry. We investigated the properties that this field satisfies
for  microscopic models of an arbitrary Kahler potential, 
in the presence of matter fields. We also investigated the decoupling of the massive
scalars (sgoldstino and matter scalars) and the effect of this on the properties of
$X$. The study can also be relevant for identifying the offshell 
couplings of the goldstino/sgoldstino to matter fields by using an effective 
approach to couple their corresponding non-linear superfields.

As it is known, in the absence of matter superfields, the action with  
$K=X^\dagger X$ and an effective
superpotential $W=f X$ and a constraint $X^2=0$, provide a superfield 
description of the Akulov-Volkov action for the goldstino fermion and
a non-linear realization of supersymmetry. This constraint essentially 
integrates (projects) out the scalar partner of goldstino, the sgoldstino. 
Such scenario can be realized in an effective model in which $K$ has 
higher dimensional Kahler terms which provide a 
mass term for  the sgoldstino; this can then be integrated out via the 
eqs of motion, to become a goldstino composite and enforce the condition $X^2=0$.
An example in this direction is provided at one-loop level
 by the familiar O'Raifeartaigh model of spontaneous supersymmetry breaking, 
with small  supersymmetry breaking, upon integrating out the other 
two massive superfields.

In the presence of additional matter fields beyond the goldstino superfield, 
the situation is more subtle and was analyzed in detail in this work.
Offshell the scalar component of $X$, $\phi_X$, is now  a mixture of the  
scalars present in the theory,  sgoldstino and scalar matter field (squark, slepton).
Using the general form of $X$ for an arbitrary $K$ and a linear superpotential
in the goldstino superfield, we computed the form of $X$ after integrating out
the scalar degrees of freedom (sgoldstino and scalar matter fields), 
which had acquired mass via the higher dimensional Kahler terms.
As a result of this, for the two-fields case 
$X$ has the property that offshell  $X^3=0$ while $X^2\not=0$. 
Thus the previous conjecture in the literature $X^2=0$
used as a condition to identify the sgoldstino has a restricted validity.
It was also shown that the sgoldstino and scalar matter field
have expressions that  are such that their  corresponding non-linear 
superfields  satisfy offshell higher order polynomial  constraints, such as 
cubic  conditions $\Phi_1^3=\Phi_1^2\,\Phi_2=\Phi_1\,\Phi_2^2=\Phi_2^3=0$ where 
$\Phi_1$ ($\Phi_2$) denote the goldstino (matter) superfields, respectively.
The offshell expressions of $\Phi_{1,2}$ can be used to couple these 
non-linear multiplets to matter. The cubic conditions shown above
can change in the case of more complicated superpotentials such as the 
R-parity violating ones or in the presence of more superfields.
The cubic conditions  of $\Phi_{1,2}$ mentioned 
are also  more general because, unlike the field $X$ and its properties, 
are independent of the choice for the improvement term.

For a better understanding of our results, 
we examined the consequences of imposing the superfield constraints $X^2=X\,\Phi_j=0$  
with $\Phi_j$ a matter superfield; these constraints were in the past 
conjectured to project out
the sgoldstino and the scalar component of $\Phi_j$, which are however
 implicitly assumed
to be infinitely massive. As a result, while $\Phi_{1,2}$ do satisfy 
offshell cubic conditions
as those mentioned above, their exact form is different from the general 
case discussed above,
due to different values of some coefficients $c_{1,2}$. 
Therefore the results derived from the superfield constraints
have a restricted validity. For two fields case
correct onshell results are possible 
when the curvature tensor $R^{ij}_{kl}$ of the Kahler 
manifold satisfies the condition $c_1\sim \det R^{1j}_{2k}=0$,  with 
$\det R^{1j}_{1k}\not =0$.
This may be possible if the action has a 
discrete R-symmetry under which the superfields have different $R$-charges.
One should keep however in mind that
it is actually the {\it offshell} constraint that is
relevant when coupling the goldstino
field to matter, in a ``non-linear'' superfield language. 

\section{Acknowledgements}

This work was supported in part by the European Commission under
ERC Advanced Grant 226371 and contract PITN-GA-2009-237920.
The work of ED was also partially supported under grant 
ANR TAPDMS ANR-09-JCJC-0146.
The authors thank G.~von~Gersdorff for many discussions  on this topic.

\end{document}